\documentclass[pra,twocolumn,amsmath,amssymb,letterpaper,superscriptaddress,nofootinbib,longbibliography]{revtex4-2}
\usepackage{graphicx}
\usepackage{dcolumn}
\usepackage{float}
\usepackage{booktabs}
\usepackage[hypertexnames, breaklinks=true, bookmarksnumbered=true, bookmarksopen=true, colorlinks=true, linktocpage=true, citecolor=blue, urlcolor=magenta, linkcolor=magenta]{hyperref}
\usepackage{amsmath,amssymb,amsfonts,amsthm}
\usepackage{mathtools}
\usepackage[T1]{fontenc}
\usepackage[latin9]{inputenc}
\usepackage{txfonts}
\usepackage{bbm}
\usepackage{bm}
\usepackage{mathrsfs}
\usepackage{xcolor}
\usepackage{natbib}
\usepackage{siunitx}
\usepackage{appendix}
\usepackage{makecell}

\begin{document}
	
	\title{Symmetry-induced fragmentation and dissipative time crystal}
	
	\author{Haowei Li}
	\affiliation{Laboratory of Quantum Information, University of Science and Technology of China, Hefei 230026, China}
	
	\author{Wei Yi}
	\email{wyiz@ustc.edu.cn}
	\affiliation{Laboratory of Quantum Information, University of Science and Technology of China, Hefei 230026, China}
	\affiliation{Anhui Province Key Laboratory of Quantum Network, University of Science and Technology of China, Hefei 230026, China}
	\affiliation{CAS Center For Excellence in Quantum Information and Quantum Physics, Hefei 230026, China}
	\affiliation{Hefei National Laboratory, University of Science and Technology of China, Hefei 230088, China}
	\affiliation{Anhui Center for Fundamental Sciences in Theoretical Physics, University of Science and Technology of China, Hefei 230026, China}
	
	\begin{abstract}
		Time crystals are a peculiar state of matter. Their emergence hinges on ergodicity breaking, which typically originates from many-body localization or Floquet prethermalization.
		Here we propose a novel scheme for devising robust dissipative time crystals where the ergodicity is broken through symmetry-induced fragmentation.
		Building upon a $U(1)$-symmetry-induced Liouville-space fragmentation, we first propose a generic Liouvillian with long-time oscillations typical of time crystals.
		We then show that, even when the $U(1)$ symmetry is broken, a prethermal time-crystal behavior survives, with distinct oscillation frequencies at different times of the steady-state approaching dynamics.
		Intriguingly, the stage-wise prethermal dynamics derive from Fermi statistics and the Liouvillian skin effect of our model---as the excitations above the boundary-localized dark states can be mapped to the irreducible representations of the permutation group, 
		the branching rules of the permutation group ensure the robustness of the prethermal time crystal.
		Our work paves the way for devising time crystals through Hilbert-space fragmentation. It also sheds light on the dynamic effects of non-Hermitian physics in many-body quantum open systems.
	\end{abstract}
	
	\maketitle
	
	\section{Introduction}
	Time crystals possess peculiar spatiotemporal structures that break the time-translation symmetry, exhibiting self-organized time-periodic dynamics~\cite{Wilczek2012,Sacha2018,Zaletel2023,Else2020}.
	A key element in devising time crystals is the breaking of ergodicity.
	For instance, in discrete time crystals~\cite{Zaletel2023,Else2020,Khemani2016,Else2016,Choi2017,Randall2021,Zhang2017,Frey2022,Mi2022,Liu2023,Else2017,Zeng2017,Kyprianidis2021,Machado2023,Yue2023,Vu2023,Pizzi2020,Deng2023,Gong2018,Chinzei2020,Yang2021,Hu2023,Xu2023,Gambetta2019}, where the discrete time-translation symmetry of a periodically driven system (or Floquet system) is spontaneously broken, the system breaks ergodicity either by resorting to the many-body localization~\cite{Khemani2016,Else2016,Choi2017,Randall2021,Zhang2017,Frey2022,Mi2022,Liu2023}, or, in an approximate fashion, through the Floquet prethermalization~\cite{Else2017,Zeng2017,Kyprianidis2021,Machado2023,Yue2023,Vu2023}.
	In the latter case, subharmonic oscillations occur within an intermediate prethermal time window, before the system finally gives in to the Floquet heating.
	In continuous time crystals that arise in the non-equilibrium dynamics of dissipative system, the ergodicity is broken through stable limit cycles~\cite{Kessler2019,Kongkhambut2022,Mattes2023,Wu2024}, emergent symmetries~\cite{Iemini2018,Piccitto2021,Prazeres2021,Passarelli2022,Cabot2023,Buca2019,Booker2020,Buca2022,Bull2022}, or strong measurements~\cite{Krishna2023}.
	But ergodicity of a quantum many-body system can also be broken through other exotic mechanisms, including the quantum many-body scars~\cite{Turner2018,Kessler2020,Serbyn2021,Papic2022,Moudgalya2022,Chandran2023,Hummel2023} and Hilbert-space fragmentation~\cite{Khemani2020,Sala2020,Yang2020,Langlett2021,Surace2022,Yoshinaga2022,Francica2023,Chattopadhyay2023}.
	Whether time-crystalline order can be stabilized under these novel circumstances is an important but open question.
	In this work, we show that robust dissipative time crystals can emerge in quantum open systems through a symmetry-induced Liouville-space fragmentation.
	We start from a general framework where the $U(1)$ symmetry of a designed Liouvillian gives rise to fragmentation, thus breaking ergodicity.
	As a result, the Liouvillian develops a series of non-dissipative eigenmodes with purely imaginary eigenvalues that are responsible for persistent oscillations at long times.
	We then focus on a dissipative lattice model of fermions as an illustrating example, and further demonstrate that
	the system is survived by prethermal time-crystal dynamics under symmetry-breaking perturbations.
	In particular, with the onset of perturbations, the eigenvalue of each non-dissipative eigenmode acquires a small real part (the Liouvillian gap of the eigenmode), whose magnitude depends on the imaginary component of the eigenvalue.
	This gives rise to a rich structure in the prethermal time-crystal dynamics---the system oscillates with distinct frequencies at different time scales of the steady-state approaching dynamics, as different eigenmodes take turns to be dominant.
	Detailed analysis reveals that both the robustness and structure of the prethermal time crystal
	are predicated by the Liouvillian skin effect and Fermi statistics.
	Under the Liouvillian skin effect~\cite{Yao2018,Kunst2018,MartinezAlvarez2018,YangFang2020,Zhang2022,Cavityskin,Haga2021,Yang2022,Hamanaka2023,Wang2023,Song2019}, the dark states of the quantum jump operators become boundary localized, forming a real-space Fermi sea.
	Fermion excitations above the dark states can then be mapped to the irreducible representations of the permutation group.
	Using the branching rules of the permutation group, we show that
	the non-dissipative eigenmodes become weakly dissipative, as they acquire eigenvalues that exhibit higher-order power-law scalings with the symmetry-breaking perturbation.
	Hence, it is this algebraic structure that protects the prethermal time-crystal behavior.
	Our work illustrates, for the first time, that symmetry-induced fragmentation, combined with the Liouvillian skin effect, can give rise to robust  time-crystal dynamics in quantum open systems.
	
	\section{General framework}
	We start from a general Lindblad master equation (setting $\hbar=1$)
	\begin{align}
		\frac{d {\rho}}{d t}=\mathcal{L}[\rho]=-\mathrm{i}[\hat{H}, {\rho}]+ \sum_\mu\gamma_\mu\left(2 \hat{K}_\mu {\rho} \hat{K}_\mu^{\dagger}-\left\{\hat{K}_\mu^{\dagger} \hat{K}_\mu, {\rho}\right\}\right),
	\end{align}
	where $\rho$ is the density matrix of the open system, $\mathcal{L}$ is the Liouvillian superoperator, $\hat{H}$ is the coherent Hamiltonian, and $\hat{K}_\mu$ are a set of quantum jump operators.
	We focus on the case where $\mathcal{L}$ possesses a $U(1)$ weak symmetry~\cite{Buca2012,liangjiang2014,Lieu2020,McDonald2022}, defined as $[\mathcal{L},\mathcal{U}^{\hat{A}}]=0$ for the symmetry superoperator $\mathcal{U}^{\hat{A}}$, where
	$\mathcal{U}^{\hat{A}}[\rho]=e^{i\theta\hat{A}}\rho e^{-i\theta\hat{A}}$ for $\theta\in [0,2\pi)$ and a Hermitian operator $\hat{A}$. Note this is a natural extension of the $Z_2$ symmetry discussed in Ref.~\cite{Minganti2018}, where $\theta$ takes only discrete values of $\{0,\pi\}$.
	It follows that $\mathcal{L}$ and $\mathcal{U}^{\hat{A}}$ can be simultaneously diagonalized, which is more transparent by vectorizing the density matrix $\rho$ and rewriting the Liouvillian as:
	$\mathcal{L}=-i\left[\hat{H}\otimes\mathbb{I}-\mathbb{I}\otimes \hat{H}^T+\sum_\mu\gamma_\mu(2\hat{K}_\mu\otimes \hat{K}^*_\mu-\hat{K}^\dagger_\mu\hat{K}_\mu\otimes\mathbb{I}-\mathbb{I}\otimes\hat{K}^T_\mu\hat{K}^*_\mu)\right]$.
	In this enlarged Hilbert space, the Liouvillian $\mathcal{L}$ can be cast into a block-diagonal matrix, where the diagonal blocks correspond to symmetry sectors with distinct eigenvalues of $\mathcal{U}^{\hat{A}}$. The Liouville space is thus fragmented under the $U(1)$ symmetry, and dynamics within a given symmetry sector is decoupled from others, breaking the ergodicity.
	
	\begin{figure}[tbp]
		\includegraphics[width=0.49\textwidth]{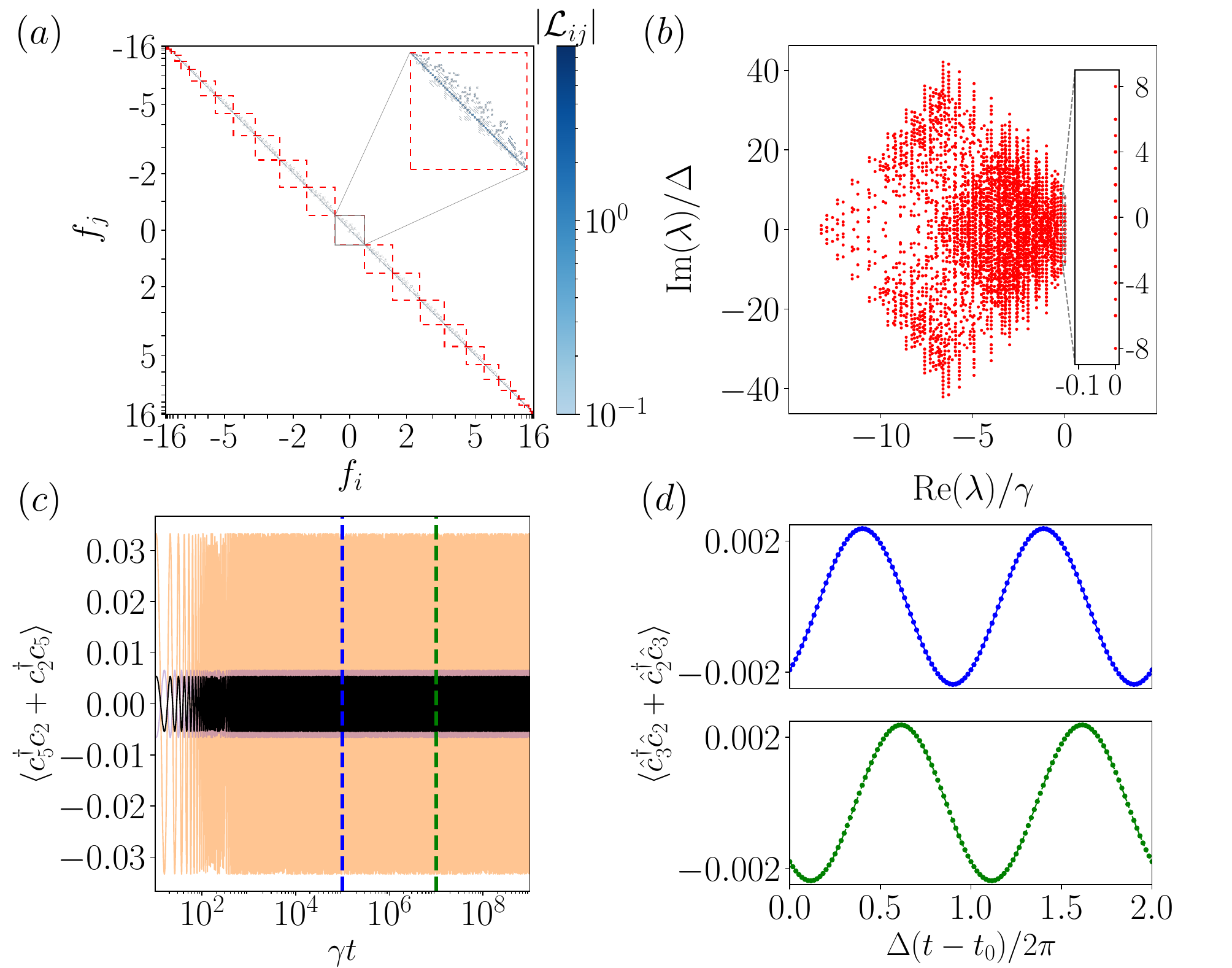}
		\caption{Symmetry-induced fragmentation and time crystal.
			(a) Block-diagonalized Liouvillian matrix. The red boxes indicate the diagonal symmetry sectors with $f_i=f_j$, the inset shows a zoomed-in view of the matrix elements (tinged according to the color bar) within a symmetry sector.
			(b) The Liouvillian spectrum on the complex plane. The inset shows the non-dissipative eigenmodes on the imaginary axis.
			(c) The long-time dynamics of $\langle \hat{c}_5^\dagger \hat{c}_2+ \hat{c}_2^\dagger \hat{c}_5\rangle$ (black: $L=10, N=4$; purple: $L=8, N=4$; orange: $L=6, N=3$).
			(d) Two periods of the oscillatory dynamics starting at $t_0\gamma=10^5$ (upper panel) and $t_0\gamma=10^7$ (lower panel), respectively marked by the blue and green vertical lines in (c).
			 Unless otherwise specified, all panels are obtained with $s=0$, $\Delta/\gamma=0.2$, $L=8$, and $N=4$.
		} \label{Fig1}
	\end{figure}
	
	We now modify the coherent Hamiltonian $\hat{H}'=\hat{H}+\Delta \hat{A}$, where $\Delta$ is a real parameter, leading to a new Liouvillian $\mathcal{L}'$.
	Apparently, $\mathcal{L}'$ has the same $U(1)$ symmetry, satisfying $[\mathcal{L}',\mathcal{U}^{\hat{A}}]=0$. Nevertheless, the Liouvillian spectrum and steady-state degeneracy are significantly modified under $\mathcal{L}'$.
	Consider a steady state $\rho^a$ of $\mathcal{L}$ that is also an eigenstate of $\mathcal{U}^{\hat{A}}$. We then have $\mathcal{U}^{\hat{A}}[\rho^a]=e^{i\theta a}\rho^a$, where $a$ denotes the difference between two eigenvalues of $\hat{A}$. Since $\mathcal{U}^{\hat{A}}[\rho^a]=\rho^a+i\theta[A,\rho^a]+\frac{(i\theta)^2}{2!}[A,[A,\rho^a]]+\cdots$ by definition, we have $[\hat{A},\rho^a]=a \rho^a$, which leads to $\mathcal{L}'[\rho^a]=-i\Delta a \rho^a$. Hence, any $\rho^a$ with $a \ne 0$ is driven out of the steady-state subspace of $\mathcal{L}'$, acquiring a purely imaginary eigenvalue $-i\Delta a$.
	Typically, the steady states of $\mathcal{L}$ can have support on multiple symmetry sectors (or consist of eigenmodes with different $a$).
	Thus, a series of eigenmodes would emerge on the imaginary axis of the Liouvillian spectrum of $\mathcal{L}'$, which reduce the steady-state degeneracy, suggesting a steady-state phase transition~\cite{Minganti2018}.
	More importantly, these eigenmodes are non-dissipative and give rise to persistent oscillations at long times, breaking the time-translation symmetry.
	Here some remarks are in order. If the eigenvalues of $\hat{A}$ are commensurate, one can always rescale $\hat{A}$, leading to integer eigenvalues. In this case, the long-time oscillations can be characterized by a single frequency $\Delta/2\pi$, corresponding to a time-crystalline order.
	Otherwise, when the eigenvalues of $\hat{A}$ are non-commensurate, the long-time oscillations consist of multiple non-commensurate frequencies, corresponding to a quasi time crystals.
	Thus, dissipative (quasi) time crystals can be engineered based on the symmetry-induced fragmentation of the Liouville space.
	
	\section{Symmetry-induced dissipative time crystal}
	To illustrate the general framework above, we consider, as an example, a one-dimensional dissipative lattice model of fermions, with the Liouvillian
	\begin{align}
		\mathcal{L}[\rho]&=-\mathrm{i}[\hat{H}, {\rho}]+ \gamma\left(2 \hat{K} {\rho} \hat{K}^{\dagger}-\left\{\hat{K}^{\dagger} \hat{K}, {\rho}\right\}\right),\label{model1}\\
		\hat{H}&=\gamma \hat{K}^\dagger \hat{K}+\Delta \hat{B}.\label{model2}
	\end{align}
	Here $\hat{c}^{\dag}_j$ ($\hat{c}_j$) is the fermion creation (annihilation) operator on site $j$,
	$\hat{B}=\sum_{j=1}^{L}  j\hat{c}_j^\dagger \hat{c}_{j}$, $L$ is the total number of lattice sites, and the jump operator $\hat{K}=\sum_{j=1}^{L-1}  \hat{c}_j^\dagger \hat{c}_{j+1}$ collectively shifts the atoms leftward (toward larger site index).
	The model can be implemented in an atom-cavity hybrid configuration, and, importantly, exhibits the Liouvillian skin effect~\cite{Song2019,Haga2021,Yang2022,Hamanaka2023,Wang2023,Cavityskin}---the geometry of the Liouvillian spectrum on the complex plane depends on the boundary condition, and the steady states are localized toward the left boundary as opposed to evenly distributed under the periodic boundary condition~\cite{Cavityskin}.
	For the remainder of this work, we only consider the open boundary condition, under which
	$[\hat{K},\hat{K}^\dag] \ne 0$, and both the dark states of $\hat{K}$ and the steady states of $\mathcal{L}$ accumulate to the left boundary.
	In the light of the Liouvillian skin effect, we introduce an anti-symmetrized excitation basis in reference to the maximally localized state where all $N$ fermions occupy the left-most $N$ sites, forming a Fermi sea in the real space.
	Specifically,  starting from the maximally localized state (denoted as $|0\rangle$), the basis $|k_{N_e},\dots, k_2, k_1\rangle$ denotes a sequential excitation of $k_1$, $k_2$, up to $k_{N_e}$ ($N_e$ is the total excitation number), where the operation $k_i$ involves moving the $(N+1-i)$th fermion (from the left) $k_i$ sites toward the right. We then have $k_1\geq k_2\geq k_3 \cdots \geq k_{N_e}$ due to the Pauli exclusion, and, as a convention, we omit any vanishing $k_i$ in the notation.
	When $\Delta=0$, the Liouvillian possesses the $U(1)$ symmetry with $\mathcal{U}^{\hat{B}}[\rho]=e^{i\theta \hat{B}}\rho e^{-i\theta\hat{B}}$, and $\theta\in [0,2\pi)$.
	It is straightforward to show that the excitation basis states defined above are eigenstates of $\hat{B}$, with
	$\hat{B}|k_{N_e},\dots, k_2, k_1\rangle=B|k_{N_e},\dots, k_2, k_1\rangle$, and the integer eigenvalue $B=\sum_{i=1}^{N_e}k_i+ N(N+1)/2$. Denoting the class of basis states with eigenvalue $B$ as $|\varphi^B\rangle$, we have $\mathcal{U}^{\hat{B}}[\rho^{B_1,B_2}]=e^{i\theta f}\rho^{B_1,B_2}$, where $\rho^{B_1,B_2}=|\varphi^{B_1}\rangle\langle\varphi^{B_2}|$ and $f=B_1-B_2$.
	Importantly, the Liouvillian $\mathcal{L}$ is block-diagonal in the basis of vectorized $\{\rho^{B_1,B_2}\}$, with each diagonal block labeled by $f$ [see Fig.~\ref{Fig1}(a)],
	and characterized by the corresponding eigenvalue $e^{i\theta f}$ of $\mathcal{U}^{\hat{B}}$.
	Since $f$ takes integer values in the range of $[-N(L-N),N(L-N)]$, there are altogether $2N(L-N)+1$ diagonal blocks.
	Such a symmetry-induced fragmentation also leaves its mark in the steady-state subspace. Specifically, a general steady state of $\mathcal{L}$ can be expressed in the basis of $\rho_{s}^{B_1,B_2}=|D^{B_1}\rangle\langle D^{B_2}|$, where the dark states $|D^{B_i}\rangle$ satisfy $\hat{K}|D^{B_{i}}\rangle=0$ and $\hat{B}|D^{B_{i}}\rangle=B_i|D^{B_{i}}\rangle$.
	Just as $\rho^a$ in the general framework, $\rho_{s}^{B_1,B_2}$ is a simultaneous eigenvector of $\mathcal{L}$ and $\mathcal{U}^{\hat{B}}$, with $\mathcal{L}[\rho_s^{B_1,B_2}]=0$ and $\mathcal{U}^{\hat{B}}[\rho_s^{B_1,B_2}]=e^{i\theta f}\rho_s^{B_1,B_2}$.
	In other words, a steady state is generally a superposition of eigenmodes $\rho^{B_1,B_2}_s$ from different symmetry sectors (or different diagonal blocks of $\mathcal{L}$).
	Now with $\Delta>0$, following the results in the general framework, we have $\mathcal{L}[\rho_{s}^{B_1,B_2}]=-i\Delta f\rho_{s}^{B_1,B_2}$.
	As illustrated in Fig.~\ref{Fig1}(b), since $f\in \mathbb{Z}$, a series of equidistant eigenmodes emerge along the imaginary axis, their eigenvalues being multiples of $i\Delta$.
	For $L,N\gg 1$, the number of these non-dissipative modes is exponentially large in $N$~\cite{Cavityskin}.
	These eigenmodes are responsible for the time-crystalline order, manifesting themselves as persistent oscillations with frequency $\Delta/2\pi$ in the long-time dynamics.
	This is apparent in Fig.~\ref{Fig1}(c)(d), where we plot the evolution of the correlation function $\langle \hat{c}_5^\dagger \hat{c}_2+ \hat{c}_2^\dagger \hat{c}_5\rangle$, for an initial state of an equal-weight superposition of all excitation basis states. A special case of our general framework was discussed in Ref.~\cite{arxiv20}, where the $U(1)$ symmetry is associated with the phase of cavity field.
	Note that, following the general framework, one can also devise dissipative time crystals where the symmetry generator $\hat{A}$ involves many-body interactions.
	
\section{Breaking the symmetry: prethermal time crystal}
\begin{figure}[tbp]
	\includegraphics[width=0.49\textwidth]{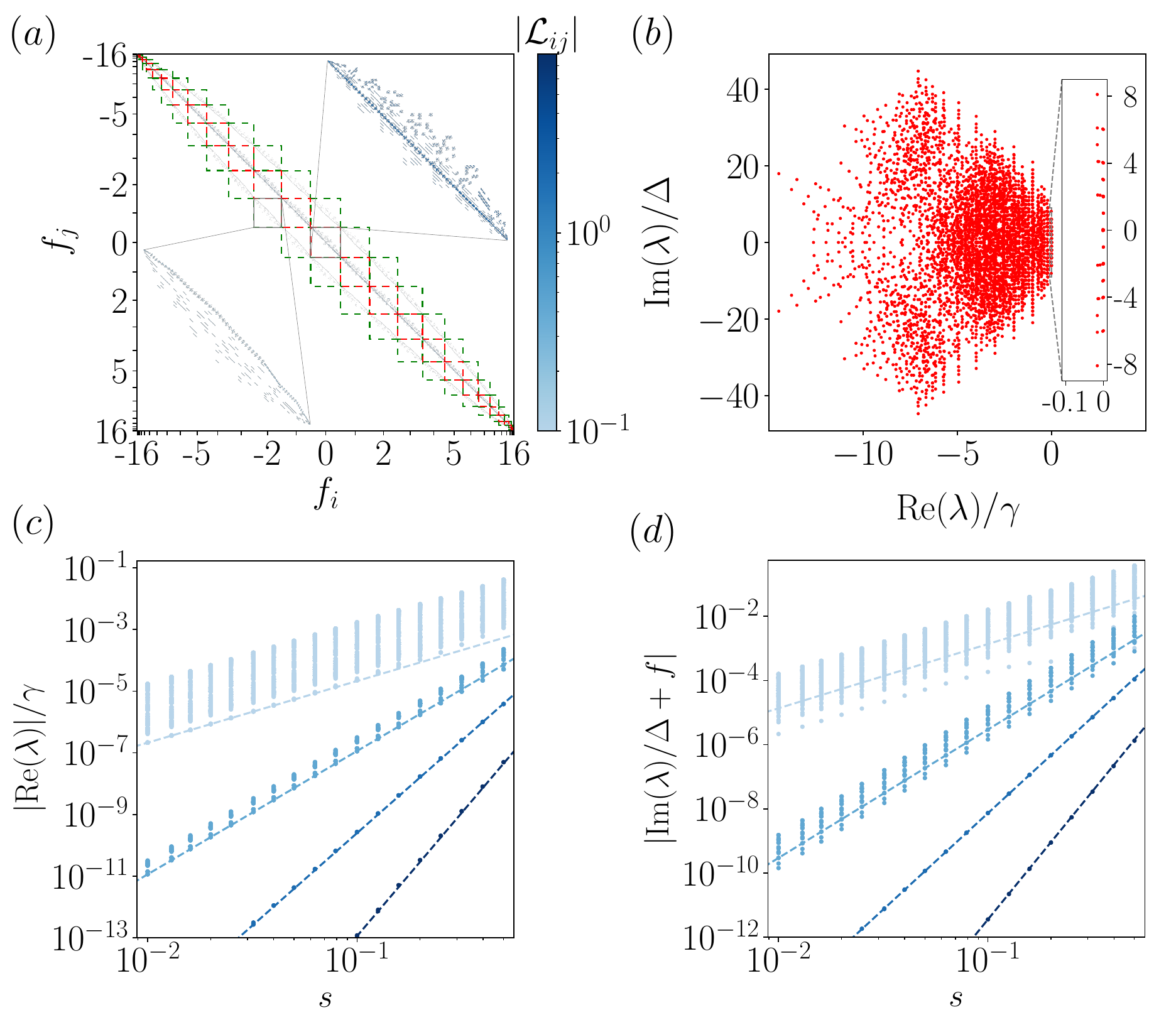}
	\caption{Prethermal time crystal. (a) Tri-diagonal Liouvillian matrix under symmetry-breaking perturbations. The red and green boxes respectively indicate the diagonal ($f_i=f_j$) and sub-diagonal ($|f_i-f_j|=1$) blocks. (b) The Liouvillian spectrum and its zoomed-in view (inset) near the imaginary axis.
		(c)(d) The real and imaginary eigenvalue components of the weakly-dissipative eigenmodes [in the inset of (b)], with respect to their corresponding unperturbed values ($-i\Delta f$).
		The dashed lines are numerical fits with the power-law scaling $\lambda_f+i\Delta f\sim s^p$, where $p=2,4,6,8$ from top to bottom. For (a)(b), we take $L=8$, $N=4$, and $s/\gamma=0.2$; while for (c)(d), $L=10$ and $N=5$. For all calculations, we set $\Delta/\gamma=0.2$.
	} \label{Fig2}
\end{figure}

An outstanding feature of time crystals is their robustness against perturbations. While symmetry is crucial for the time crystal discussed above,
it is natural to question its fate in the presence of symmetry-breaking perturbations.
For this purpose, we consider adding a symmetry-breaking term $\hat{H}_1=s(\hat{K}+\hat{K}^\dagger)$ to the coherent Hamiltonian, where $s>0$.
As a result, the Liouvillian becomes tri-block diagonal, as illustrated in Fig.~\ref{Fig2}(a). While the previously non-dissipative eigenmodes no longer stay on the imaginary axis, they do not deviate much on the complex plane even for a considerably large symmetry-breaking term [see Fig.~\ref{Fig2}(b)].
More quantitatively, we plot the real and imaginary components of the eigenvalues of the perturbed eigenmodes as functions of $s$ in Fig.~\ref{Fig2}(c) and (d), respectively.
From the numerical fit, we find that the Liouvillian eigenvalue $\lambda_f$ nearby $-i\Delta f$ satisfies a higher-order power-law scaling
\begin{align}
	\lambda_f+i\Delta f\sim s^{p},
\end{align}
with the exponent $p$ dependent on $f$, or the location of the eigenmode along the imaginary axis. While the small real components suggest the domination of the long-time dynamics by these eigenmodes, the small deviation from $-i\Delta f$ implies negligible frequency shift of the oscillations.

To understand the higher-order scaling behavior, we adopt the Liouvillian perturbation formalism developed in Ref.~\cite{Cavityskin}, and rewrite the Liouvillian as $\mathcal{L}=\mathcal{L}_0+s\mathcal{L}_1$, where $\mathcal{L}_1[\rho]=-i[\hat{K}+\hat{K}^\dagger,\rho]$.
Denoting $\rho_0=\rho_{s}^{B_1,B_2}$ as the steady state of $\mathcal{L}_0$ in the symmetry sector labeled by $f=B_1-B_2$, we further construct the density matrix in a perturbative fashion
\begin{align}
	\rho=\rho_0+\sum_{n=1}^{\infty}s^n\rho_n,\label{eq:rhoinfty}
\end{align}
where $\rho_n$ is to be determined. In the following, we first show that an exact non-dissipative eigenmode of $\mathcal{L}$ can be constructed following Eq.~(\ref{eq:rhoinfty}) in the thermodynamic limit $L,N\rightarrow \infty$. This would help us to understand the more relevant case of finite $L$ and $N$.

\begin{figure}[tbp]
	\includegraphics[width=0.5\textwidth]{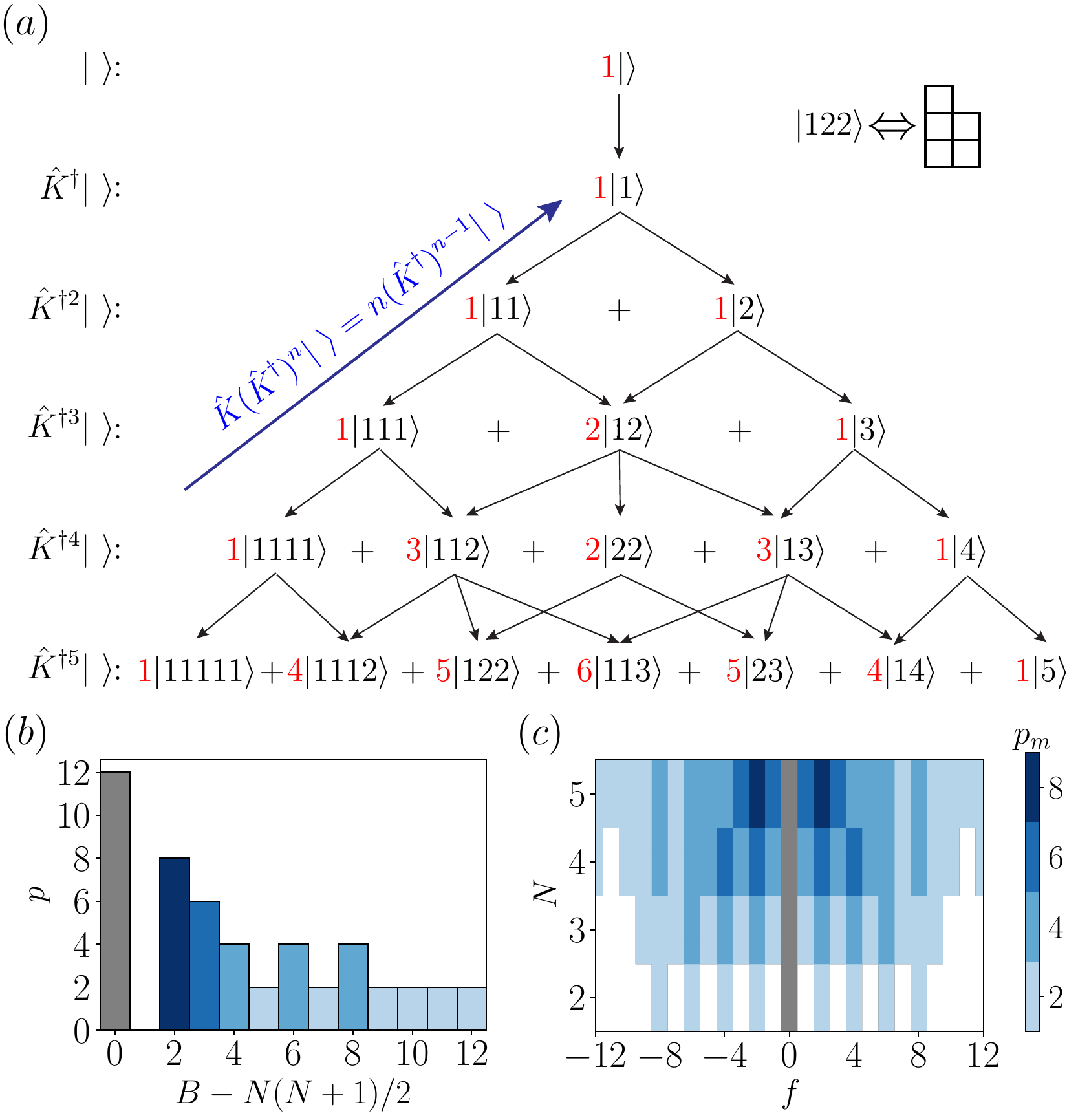}
	\caption{Mapping to the permutation group.
		(a) Excited states, in the excitation basis, generated from the dark state $|0\rangle$ by $\hat{K}^\dag$.
		The expansion coefficients (red) coincide with the dimensions of the corresponding irreducible representation (of the permutation group). An example of the Young-diagram correspondence of the
		excitation basis is given in the upper right corner, with which the structure in (a) corresponds to the Young's lattice.
		(b) Relation of the exponent $p=2l^{|D^{B}\rangle}+2$ with $B$ for $L=10$ and $N=5$, obtained analytically from the group analysis.
		(c) Maximum exponent $p_m$ for eigenmodes characterized by $f$, numerically calculated for different $N$ and $L=10$.
		For (b)(c), we take $s/\gamma=0.2$, and $\Delta/\gamma=0.2$.
	} \label{Fig3}
\end{figure}

A key observation is that each excitation basis state $|k_{N_e},\dots ,k_1\rangle$ can be mapped to an irreducible representation of the permutation group.
To see this explicitly, we start from the actions of $\hat{K}$ and $\hat{K}^\dagger$ on the excitation basis:
\begin{align}
	\hat{K}|k_{N_e},k_{N_e-1},\dots ,k_1\rangle&=\sum_{i=1}^{N_e}(1-\delta_{k_i,k_{i+1}})|p^{(i)}_{N_e},p^{(i)}_{N_e-1},\dots,p^{(i)}_1\rangle,
	\label{KFock}\\
	\hat{K}^\dagger|k_{N_e},k_{N_e-1},\dots ,k_1\rangle&=\sum_{i=1}^{N_e+1}(1-\delta_{k_{i-1},k_i})|q^{(i)}_{N_e+1},q^{(i)}_{N_e},\dots ,q^{(i)}_1\rangle,
	\label{KDFock}
\end{align}
where we set $k_{N_e+1}=0$, $p_j^{(i)}=k_j-\delta_{i,j}$, $q_j^{(i)}=k_j+\delta_{i,j}$, and require $\delta_{k_0,k_1}=0$. The Kronecker deltas enforce the Pauli exclusion, ensuring fermionic statistics.
Expanding $(\hat{K}^\dagger)^n|0\rangle$ on the excitation basis, $(\hat{K}^\dagger)^n|0\rangle=\sum_k b_{n,k}|\phi_n^k\rangle$ with $\sum_i k_i=n$, we obtain the recursion relation
\begin{align}
	b_{n,k}=\sum_{r\nearrow k}b_{n-1,r},
\end{align}
where $r\nearrow k$ denotes that $\langle\phi_{n-1}^r|\hat{K}|\phi_n^k\rangle=1$.

The same recursive structure appears in the branching rules of the permutation group (see the Appendix).
For a set of $n$ elements, an irreducible representation is labeled by a partition $[k_{N_e},k_{N_e-1},\dots,k_1]$ with $k_1\ge k_2\ge\dots\ge k_{N_e}$ and $\sum_j k_j=n$.
Each partition corresponds to a Young diagram with $k_j$ boxes in the $j$th row.
Restriction and induction of representations correspond respectively to removing and adding one box in the Young diagram~\cite{Saganmain,Vershikmain}:
\begin{align}
	\mathrm{Res}[k_{N_e},\dots ,k_1]&=\bigoplus_{i=1}^{N_e}(1-\delta_{k_i,k_{i+1}})[p^{(i)}_{N_e},p^{(i)}_{N_e-1},\dots ,p^{(i)}_1],\\
	\mathrm{Ind}[k_{N_e},\dots ,k_1]&=\bigoplus_{i=1}^{N_e+1}(1-\delta_{k_{i-1},k_i})[q^{(i)}_{N_e+1},q^{(i)}_{N_e},\dots ,q^{(i)}_1].
\end{align}
Comparing with Eqs.~(\ref{KFock})-(\ref{KDFock}), one finds that $\hat{K}$ and $\hat{K}^\dagger$ act exactly as the restriction and induction operations on the corresponding representations.
If $c_{n,k}$ denotes the dimension of the irreducible representation $[\phi_n^k]$, then from the branching rules,
\begin{equation}
	c_{n,k}=\sum_{r\nearrow k}c_{n-1,r},\qquad
	nc_{n-1,r}=\sum_{r\nearrow k}c_{n,k}.
	\label{branching}
\end{equation}
Identifying $b_{n,k}=c_{n,k}$ (which can be proved by induction from $b_{0,0}=1$), we have
\begin{equation}
	\hat{K}(\hat{K}^\dagger)^n|0\rangle
	=\sum_k b_{n,k}\hat{K}|\phi_n^k\rangle
	=n\sum_r b_{n-1,r}|\phi_{n-1}^r\rangle
	=n(\hat{K}^\dagger)^{n-1}|0\rangle,
	\label{eq:restrict}
\end{equation}
which is the key relation used in our derivation and illustrated in Fig.~\ref{Fig3}(a).

Furthermore, we observe numerically that if we replace $|0\rangle$ by other dark states $|D\rangle$ of $\hat{K}$, the relation still holds: $\hat{K}(\hat{K}^\dagger)^n|D\rangle=n(\hat{K}^\dagger)^{n-1}|D\rangle$.
Hence for any pair of dark states $|D^{B_1}\rangle$ and $|D^{B_2}\rangle$, we can set $\rho_n=\sum_{j=0}^n\alpha_n^{(j)}\hat{K}^{\dagger j}|D^{B_1}\rangle\langle D^{B_2}|\hat{K}^{n-j}$,
with coefficients
\[
\alpha_n^{(j)}=\frac{i^{n-2j}s^n}{((1-i)\gamma-i\Delta)^{n-j}((1+i)\gamma+i\Delta)^j(n-j)!j!},
\]
so that the recurrence relation
\begin{align}
	\mathcal{L}_0[\rho_{n}]+i\Delta f\rho_{n}=-s\mathcal{L}_1[\rho_{n-1}],\label{eq:recur}
\end{align}
holds for all $n\in\mathbb{N}^+$ when $L,N\to\infty$. 
Therefore $\rho=\rho_0+\sum_{n=1}^{\infty}s^n\rho_n$ is an exact non-dissipative eigenmode of $\mathcal{L}$ with eigenvalue $-i\Delta f$.
For finite $L$ or $N$, however, the mapping ceases to apply at sufficiently large $n$ due to the limited Hilbert space, and the perturbative expansion must be truncated.
Defining $l_N^{|D\rangle}=\max\{j|\langle D|\hat{c}_j^\dagger\hat{c}_j|D\rangle=1\}$ and
$l_L^{|D\rangle}=\min\{j|\langle D|\hat{c}_{L-j}^\dagger\hat{c}_{L-j}|D\rangle>0\}$,
we find that $\hat{K}(\hat{K}^\dagger)^n|D\rangle=n(\hat{K}^\dagger)^{n-1}|D\rangle$ holds only up to $n\le l^{|D\rangle}=\min(l_N^{|D\rangle},l_L^{|D\rangle})$.
Thus for $\rho_0$ being the steady state of $\mathcal{L}_0$, the truncated perturbative series $\rho=\rho_0+\sum_{n=1}^{l}s^n\rho_n$
is an approximate eigenmode of $\mathcal{L}$ with eigenvalue $\lambda_f+i\Delta f\sim s^p$, where $p=2l+2$.

	\begin{table}[tbp]
		\setlength{\tabcolsep}{2mm}
		\begin{tabular}{c|c|c|c|c}		
			\hline
			\makecell[c]{$|D_1\rangle$,$|D_2\rangle$ of the eigenmode $|D_1\rangle\langle D_2|$} &$|f|$&$N$&$(l_N^{|D_1\rangle},l_L^{|D_1\rangle})$&$p_m$\\
			\hline
			\makecell[c]{$|111\rangle-|22\rangle+|3\rangle$\\$|11\rangle-|2\rangle$}
			&1&\makecell[c]{$3$\\$4$\\$5$}&\makecell[c]{$(0,4)$\\$(1,3)$\\$(2,2)$}&\makecell[c]{$2$\\$4$\\$6$}\\
			\hline
			\makecell[c]{$|11\rangle-|2\rangle$\\$|0\rangle$}
			&2&\makecell[c]{$2$\\$3$\\$4$\\$5$}&\makecell[c]{$(0,6)$\\$(1,5)$\\$(2,4)$\\$(3,3)$}&\makecell[c]{$2$\\$4$\\$6$\\$8$}\\
			\hline
			\makecell[c]{$|111\rangle-|12\rangle+|3\rangle$\\$|0\rangle$}
			&3&\makecell[c]{$3$\\$4$\\$5$}&\makecell[c]{$(0,4)$\\$(1,3)$\\$(2,2)$}&\makecell[c]{$2$\\$4$\\$6$}\\
			\hline
			\makecell[c]{$|22\rangle-|13\rangle+|4\rangle$\\$|0\rangle$}
			&4&\makecell[c]{$2$\\$3$\\$4$\\$5$}&\makecell[c]{$(0,4)$\\$(1,3)$\\$(2,2)$\\$(3,1)$}&\makecell[c]{$2$\\$4$\\$6$\\$4$}\\
			\hline
			\makecell[c]{$|122\rangle-|113\rangle+|23\rangle+2|14\rangle-2|5\rangle$\\$|0\rangle$}
			&5&\makecell[c]{$3$\\$4$}&\makecell[c]{$(0,2)$\\$(1,1)$}&\makecell[c]{$2$\\$4$}\\
			\hline
			\makecell[c]{$|44\rangle-|134\rangle+|224\rangle+|1133\rangle-|1223\rangle$\\$ |111\rangle-|12\rangle+|3\rangle$}&5&\makecell[c]{$5$}&\makecell[c]{$(1,1)$}&\makecell[c]{$4$}\\
			\hline
			\makecell[c]{$|33\rangle-|24\rangle+|15\rangle-|6\rangle$\\$|0\rangle$}
			&6&\makecell[c]{$2$\\$3$}&\makecell[c]{$(0,2)$\\$(1,1)$}&\makecell[c]{$2$\\$4$}\\
			\hline
			\makecell[c]{$|123\rangle-|114\rangle+|24\rangle-|222\rangle-2|33\rangle$\\$|0\rangle$}
			&6&\makecell[c]{$4$\\$5$}&\makecell[c]{$(1,2)$\\$(2,1)$}&\makecell[c]{$4$\\$4$}\\
			\hline
		\end{tabular}
		\caption{Analytic results from the permutation-group analysis, for $L=10$ and $|f|\leq 6$.}
		\label{table1}
	\end{table}
	
	\section{Prethermal dynamics}
	
	The power-law scaling of the eigenvalues suggest that these eigenmodes are weakly dissipative, which gives rise to rich prethermal dynamics.
	For instance, when $L=2N$, the steady state $\rho_s=|11\rangle\langle\,\, |-|2\rangle\langle \,\,|$ of $\mathcal{L}_0$ acquires an eigenvalue $\lambda +i2\Delta \sim s^{2N-2}$ under the symmetry-breaking perturbation (see Table.~\ref{table1}).
	Therefore, the thermalization time of the mode scales as $\sim s^{2-2N}$, increasing exponentially with $N$. This slow thermalization manifests in the dynamics as persistent oscillations at the frequency
	$\Delta/\pi$ in the thermodynamic limit, and is not just a transient oscillatory effect. Thus, depending on the initial state, a prethermal time-crystalline behavior emerges, characterized by distinct oscillatory frequencies appearing at different times during the approach to the steady state.
	In Fig.~\ref{Fig3}(b), we show the analytically calculated exponent $p$ for all dark states in a chain of $L=10$ at half filling $N=5$.
	Generally, $p$ increases with decreasing $B$, meaning localized states (with smaller $B$) tend to be more robust dynamically, as they feature a smaller Liouvillian gap $\text{Re}\lambda_f$.
	On the other hand, for a given $f$, there can be many weakly dissipative eigenmodes with different exponents $p$. We thus define $p_m$ as the maximum among them. In Fig.~\ref{Fig3}(c), we plot $p_m$ as functions of $f$ and $N$ for a lattice with $L=10$, which are obtained by numerically fitting the Liouvillian spectrum.
	The numerically calculated $p_m$ (and the corresponding eigenmodes) are consistent with the analytical results, tabulated in Table.~\ref{table1}.
	Generally, for a fixed $N$ (fixed $f$), $p_m$ becomes larger for smaller $f$ (larger $N$).
	This means that: i) in a system with fixed atom number, oscillations with smaller frequencies
	dominate at later times; ii) with increasing atom number, oscillations with the same frequency emerge later.
	
\begin{figure}[tbp]
		\includegraphics[width=0.5\textwidth]{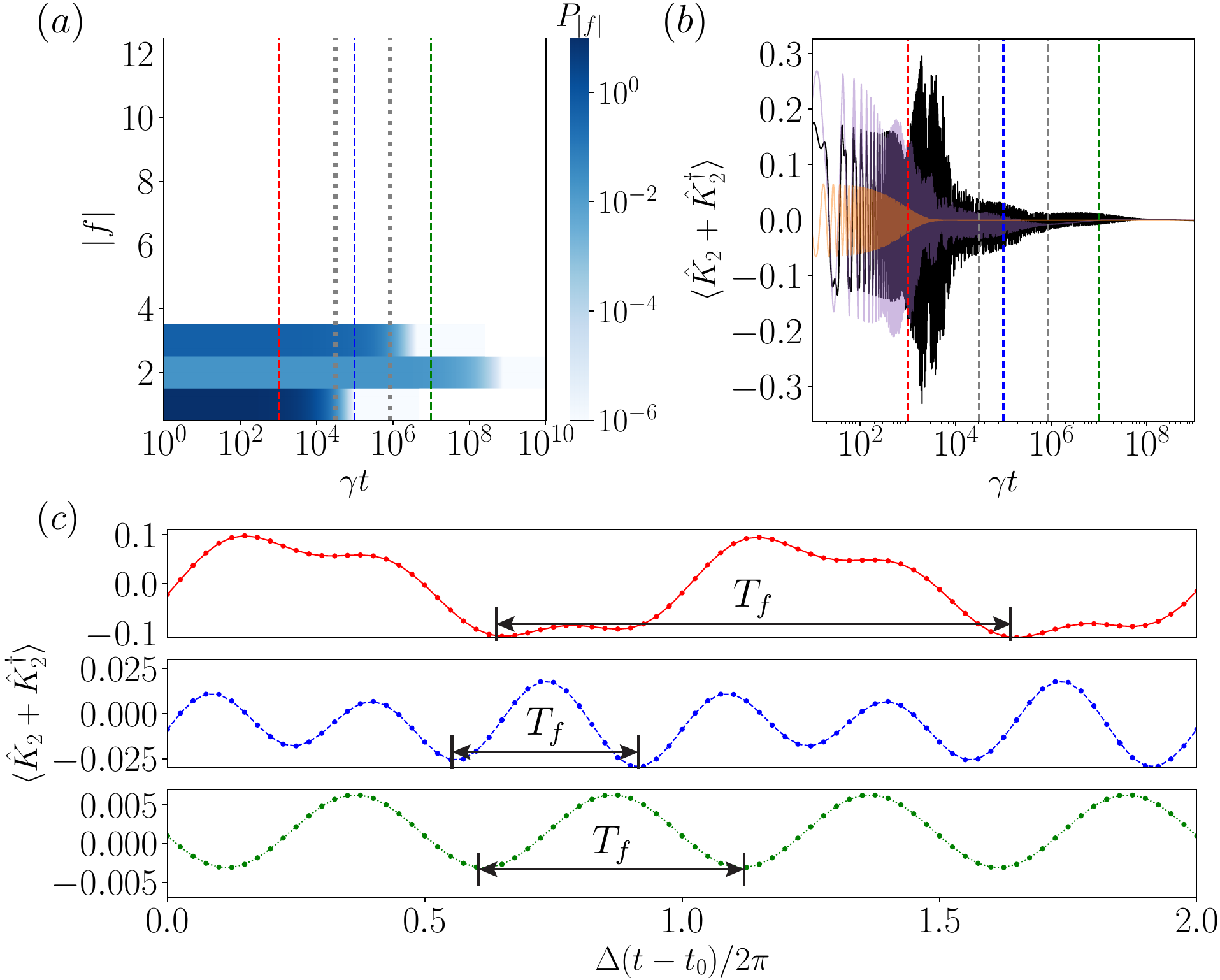}
		\caption{Stage-wise dynamics of the prethermal time crystal.
			(a) Time evolution of $P_{|f|}$.
			(b) Numerically simulated evolution of $\langle\hat{K}_2+\hat{K}^\dagger_2\rangle$, where  $\hat{K}_2=\sum_{j=1}^{L-2}  \hat{c}_j^\dagger \hat{c}_{j+2}$ (black: $L=10, N=4$; purple: $L=8, N=4$; orange: $L=6, N=3$).
			In (a)(b), stages with different dominant eigenmodes are separated by gray vertical lines.
			(c) Oscillations zoomed in, starting at different times $t_0\gamma=10^3$ (upper), $t_0\gamma=10^5$ (middle) and $t_0\gamma=10^7$ (lower), which are labeled with vertical lines in (a)(b) with the corresponding colors.
			The oscillation period is given by $T_f=2\pi/f\Delta$, with $f=1,3,2$, respectively, for upper, middle and lower panels, consistent with (a).
			 All panels are obtained with $s/\gamma=0.2$, and $\Delta/\gamma=0.2$, and we take $L=10$, $N=4$ in (a) and (c).
		} \label{Fig4}
	\end{figure}

	We numerically confirm the above conclusions by simulating the open-system
	dynamics for $L=10$ and $N=4$, as shown in Fig.~\ref{Fig4}.
	We initialize the system such that the initial density matrix has support on
	eigenmodes with $f=\pm 1$ ($p_m=2$), $f=\pm 2$ ($p_m=6$), and $f=\pm 3$ ($p_m=4$).
	The overlap of the time-dependent density matrix with eigenmodes in the vicinity
	of the eigenvalue $-i\Delta f$ is defined as
	$P_{|f|}=\sum_{|B_1-B_2|=|f|}\mathrm{Tr}(\rho\,\rho_{s}^{B_1,B_2})$.
	With different values of $p_m$, different eigenmode sectors dominate the
	dynamics at different times, giving rise to the stage-wise oscillatory behavior
	observed in Fig.~\ref{Fig4}(a,b).
	The oscillations on distinct time scales exhibit characteristic frequencies
	dictated by $f$ [see Fig.~\ref{Fig4}(c)], in agreement with the theoretical
	predictions summarized in Table~\ref{table1}.
		In addition, Fig.~\ref{Fig4}(b) includes results for different system sizes
		(purple: $L=8, N=4$; orange: $L=6, N=3$),
		which reveal that the thermalization time increases exponentially with the
		system size, indicating a pronounced finite-size dependence of the relaxation
		dynamics.

	\section{Discussion}
	We show that robust time crystals can be engineered based on a $U(1)$-symmetry-induced fragmentation,
	wherein structured prethermal time-crystal dynamics persist under symmetry-breaking perturbations.
	The robustness derives from the Liouvillian skin effect and Fermi statistics, under whose impact an elegant permutation-group representation of the excitations exists based on the limited available Hilbert space above the real-space Fermi sea.
	The dissipative time crystals thus formed are distinct from those under the
	dynamical symmetries~\cite{Buca2019,Booker2020,Buca2022}, wherein the Liouville space is not fragmented and the ergodicity is broken for a different mechanism (see the Appendix).
	Our work highlights the pivotal role of non-Hermitian phenomena, such as the Liouvillian skin effect, in the dynamics of many-body quantum open systems.
	
	\begin{acknowledgments}
		We thank Heran Wang, Zhong Wang, and Berislav Bu\v{c}a for helpful discussions.
		This work is supported by the National Natural Science Foundation of China (Grant No. 12374479), and by the Innovation Program for Quantum Science and Technology (Grant No. 2021ZD0301205).
	\end{acknowledgments}
	
	\begin{appendix}
		
		\section{An alternative example of symmetry-induced fragmentation and time crystal}
\begin{figure*}[tbp]
			\includegraphics[width=0.95\textwidth]{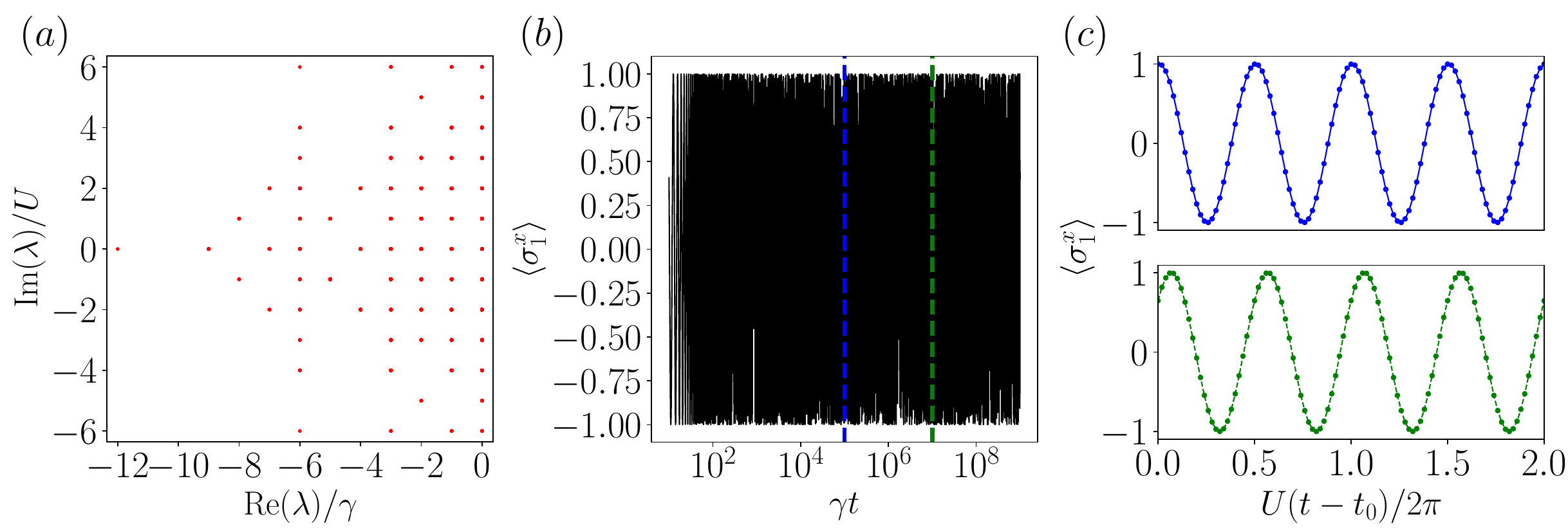}
			\caption{(a) Liouvillian spectrum for Eq.~(\ref{Lioudoublon}) with $N=6$. (b) The long-time dynamics of $\langle\sigma_1^x\rangle$. (c) A zoomed-in view of (b) over four periods, with different starting times of $t_0\gamma=10^5$ (blue) and $t_0\gamma=10^7$ (green), respectively, as marked by the vertical dashed lines in (b). For all simulations, we take $U=\gamma$.
			} \label{Fig5}
		\end{figure*}

		In the main text, we devise a dissipative time crystal based on the $U(1)$ symmetry with a single-particle symmetry generator. In this Appendix we illustrate a more involved case where the symmetry generator features many-body interactions.
		We consider the following Liouvillian in a one-dimensional spin-$1/2$ chain under the periodic boundary condition (setting $\hbar=1$)
		\begin{align}
			\frac{d {\rho}}{d t}=\mathcal{L}[\rho]=-\mathrm{i}[\hat{H}, {\rho}]+ \sum_{j=1}^N\gamma\left(2 \hat{L}_j {\rho} \hat{L}^{\dagger}_j-\left\{\hat{L}^{\dagger}_j \hat{L}_j, {\rho}\right\}\right),
			\label{Lioudoublon}
		\end{align}
		where the coherent Hamiltonian and jump operators read
		\begin{align}
			\hat{H}=\frac{U}{4}\sum_j (\sigma^z_j+1)(\sigma^z_{j+1}+1),\quad \hat{L}_j=\frac{\gamma}{4} (1-\sigma^z_{j-1})\sigma^+_j(1-\sigma^z_{j+1}).
			\label{Hdoublon}
		\end{align}
		Here $j$ indicates the site index of the spin, $N$ is the lattice size, and $U$ and $\gamma$ are the interaction and dissipation rates, respectively.
		The Liouvillian (\ref{Lioudoublon}) has the following $U(1)$ symmetry
		\begin{align}
			[\mathcal{L},\mathcal{U}^{\hat{H}}]=0,\quad\text{with}\quad \mathcal{U}^{\hat{H}}[\rho]=e^{i\theta \hat{H}} \rho e^{-i\theta \hat{H}},
			\label{Symmdoublon}
		\end{align}
		where $\theta\in [0,2\pi)$, and $\hat{H}$ also plays the role of the symmetry generator.
		The eigenstates of $\hat{H}$ are product states of spin-up $\left|\uparrow\right\rangle$ and spin-down $\left|\downarrow\right\rangle$ states along the chain. For a given product state, its eigenvalue is determined by counting the number of adjacent spin-up pairs $\left|\uparrow\uparrow\right\rangle$.
		The eigenvalue thus takes integer values in the range of $[0,N]$.
		Following the derivations in the main text, a series of equidistant Liouvillian eigenmodes
		should emerge on the imaginary axis with $\lambda_n=inU$ ($n\in \mathbb{Z}$). This is illustrated in Fig.~\ref{Fig5}(a) for a chain with $N=6$.
		These non-dissipative eigenmodes give rise to persistent oscillations in the long-time dynamics. This is shown in Fig.~\ref{Fig5}(b)(c). For our numerical simulations,
		we initialize the system in a pure state, with the density matrix $\rho(t=0)=(|\phi_1\rangle\langle\phi_1|+|\phi_1\rangle\langle\phi_2|+|\phi_2\rangle\langle\phi_1|+|\phi_2\rangle\langle\phi_2|)/2$, where
		$|\phi_1\rangle=\left|\uparrow\uparrow\uparrow\uparrow\uparrow\uparrow\right\rangle$, and $|\phi_2\rangle=\left|\downarrow\uparrow\uparrow\uparrow\uparrow\uparrow\right\rangle$.
		Since $|\phi_1\rangle$ and $|\phi_2\rangle$ are eigenstates of $\mathcal{U}^{\hat{H}}$ with eigenvalues $e^{2i\theta}$ and $e^{-2i\theta}$, respectively,
		the oscillation frequency is $U/\pi$, consistent with the numerical results in Fig.~\ref{Fig5}(b)(c).
		
		\section{Implementing the tilted lattice model}
\begin{figure}[tbp]
			\includegraphics[width=0.45\textwidth]{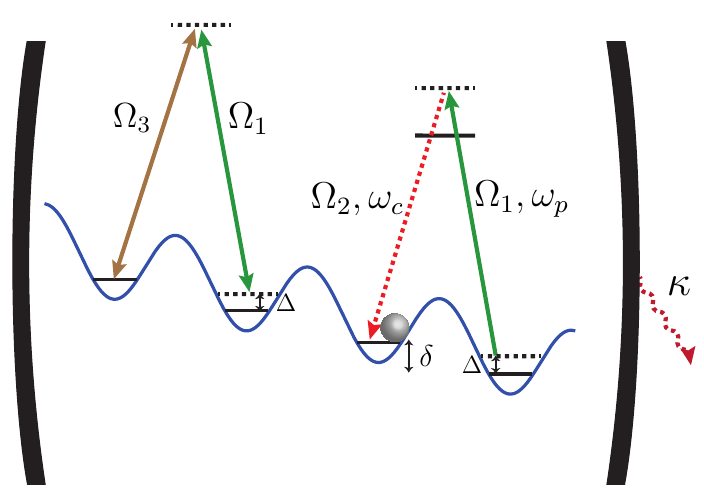}
			\caption{Schematics for implementing the tilted lattice model in a atom-cavity hybrid setup. Here $\Omega_{1,2,3}$ are the single-photon Rabi frequencies of the laser and cavity couplings, $\omega_p$ is the frequency of the pumping laser (marked in green), $\omega_c$ is the cavity-field frequency (marked in red dashed), $\Delta_c=\omega_c-\omega_p$ is the cavity detuning, and $\kappa$ is the cavity decay rate.} \label{Fig6}
		\end{figure}

		The tilted lattice model in the main text can be implemented using an atom-cavity hybrid setup, similar to the one in Ref.~\cite{Cavityskin}.
		As illustrated in Fig.~\ref{Fig6}, spinless fermions in a tilted one-dimensional lattice are coupled to a cavity. While the detuning $\delta$ between adjacent sites suppresses direct inter-site hopping, two Raman-processes are introduced to induce laser-assisted hopping. One of the Raman processes is assisted by the cavity field (red dashed arrow). Given a lossy cavity, dynamics of the fermions adiabatically follows that of the cavity field, and the fermions are effectively subject to a cavity-dependent dynamic gauge potential.
		An additional two-photon detuning $\Delta$ is introduced for both Raman processes, which is absent in
		Ref.~\cite{Cavityskin}.
		Adiabatically eliminating the cavity field and taking the tight-binding approximation, the atomic density matrix $\rho$ is governed by the Lindblad master equation (with $\hbar=1$)
		\begin{align}
			\frac{d {\rho}}{d t}=\mathcal{L}[\rho]=-\mathrm{i}[\hat{H}, {\rho}]+ \gamma\left(2 \hat{K} {\rho} \hat{K}^{\dagger}-\left\{\hat{K}^{\dagger} \hat{K}, {\rho}\right\}\right),\label{eq:slind}
		\end{align}
		with the coherent Hamiltonian
		\begin{align}
			\hat{H}=-\frac{\Delta_c}{\kappa}\gamma \hat{K}^\dagger \hat{K}+s(\hat{K}+\hat{K}^\dagger)+\Delta\hat{B}.
			\label{eq:effH}
		\end{align}
		Here $\gamma=\kappa \lambda^2/(\Delta_c^2+\kappa^2)$, the quantum jump operator $\hat{K}=\sum_{j=1}^{L-1}  \hat{c}_j^\dagger \hat{c}_{j+1}$, $\Delta_c$ is the cavity detuning, and $\hat{B}=\sum_{j=1}^{L}  j\hat{c}_j^\dagger \hat{c}_{j}$. Setting $\Delta_c/\kappa=-1$, we get the Liouvillian Eqs.~(\ref{model1})(\ref{model2}).
		
		\section{Distinction from dynamical symmetries}
		In quantum open systems, dissipative time crystals can also derive from dynamical symmetries~\cite{Buca2019,Booker2020,Buca2022}.
		Specifically, given a Lindblad master equation of the form (\ref{eq:slind}), the dynamical symmetry corresponds to the existence of an operator $\hat{A}$, satisfying $[\hat{H},\hat{A}]=\lambda \hat{A}$, and $[\hat{K},\hat{A}]=[\hat{K}^\dagger,\hat{A}]=0$. It is then straightforward to construct, from any steady state $\rho_s$, the Liouvillian eigenmodes $\rho_{nm}=\hat{A}^n\rho_s (\hat{A}^{\dag})^m$, with purely imaginary eigenvalues $-i(n-m)\lambda$. These non-dissipative modes thus give rise to long-time oscillatory dynamics. However, the relevant $\mathcal{U}^{\hat{A}}$  does not generally correspond to a weak symmetry of the system, that is $[\mathcal{L},\mathcal{U}^{\hat{A}}] \ne 0$. Consequently, the dynamical symmetries do not lead to Liouville-space fragmentation.
		By contrast, our scheme in the main text crucially relies on the weak symmetry and the symmetry-induced fragmentation to achieve ergodicity breaking. As we illustrated in the previous section, starting from Eqs.~(\ref{eq:slind})and (\ref{eq:effH}), we have $[\hat{H},\hat{B}]=0$, and $[\hat{K},\hat{B}]=\hat{K}$ in our concrete model, such that $[\mathcal{L},\mathcal{U}^{\hat{B}}]=0$. The non-dissipative eigenmodes arise for $\Delta \ne 0$, crucially dependent on the Liouville-space fragmentation and in sharp contrast to those under the dynamical symmetries.
		
		\section{Liouvillian skin effect of the tilted lattice model}
\begin{figure*}[tbp]
			\includegraphics[width=0.95\textwidth]{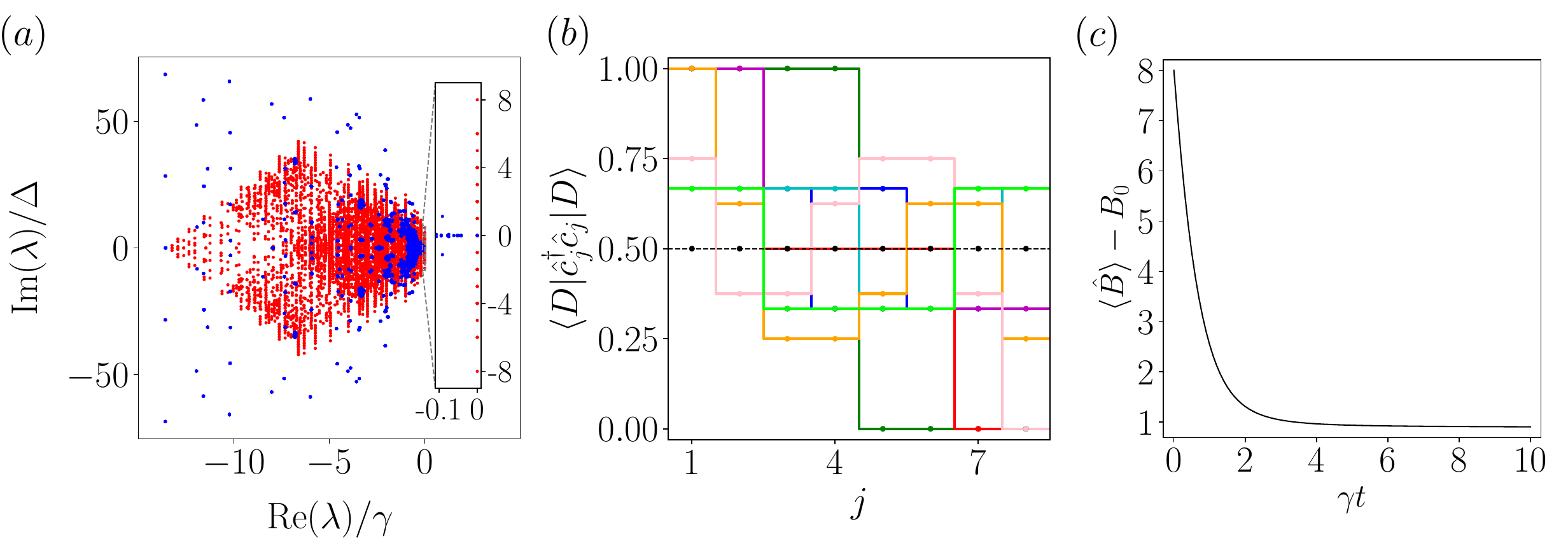}
			\caption{(a) Liouvillian spectra under the open boundary condition (red) and the periodic boundary condition (blue), respectively. (b) Spatial distribution of all the dark states under the open boundary condition (solid lines) and the periodic boundary condition (black dashed line). (c) Evolution of $\langle \hat{B}\rangle$. For all figures, we take $L=8$, $N=4$, and $\Delta/\gamma=0.2$.
			} \label{Fig7}
		\end{figure*}

		The titled lattice model Eqs.~(\ref{model1})(\ref{model2}) exhibits the Liouvillian skin effect. This is manifest in the difference of Liouvillian spectra under different boundary conditions, the boundary localization of the steady states under the open boundary condition, and atomic density dynamics.
		First, in Fig.~\ref{Fig7}(a), we show the Liouvillian spectra under both the periodic boundary condition (blue) and the open boundary condition (red).
		The eigenspectra are drastically different under different boundary conditions, which is a key signature of the Liouvillian skin effect. In particular, under the periodic boundary condition, there are no non-dissipative eigenmodes on the imaginary axis, meaning the time crystal behavior only exists under the open boundary condition.
		Second, under the open boundary condition, all dark states of $\hat{K}$ become boundary localized, as are the steady states of the Liouvillian. In Fig.~\ref{Fig7}(b), we plot the spatial distribution of the dark states (solid lines), which are all localized toward the left boundary. Whereas under the open boundary, the distribution is uniform (black dashed line).
		The Liouvillian skin effect also manifests as the boundary-approaching dynamics. In Fig.~\ref{Fig7}(c), we show the time evolution of $\langle\hat{B}\rangle-(N+1)N/2$, with the system initialized in an equal-weight superposition of all excitation basis [which is the same as Fig.~\ref{Fig1}(c)(d)]. Here $\langle\hat{B}\rangle$ is an indication of the extent of localization of the time-evolved state: the state is more localized with smaller $\langle\hat{B}\rangle$. Evidently,
		$\langle \hat{B}\rangle$ rapidly decreases toward the steady-state value in the dynamics, which also indicates the localized nature of the steady states.
		
	\section{Mapping between excitations with the permutation-group representation}
	In this appendix, we provide more details on the mapping between the excitation states and irreducible representations of the permutation group. The states generated by the action of $\hat{K}^\dag$ can be mapped onto the permutation group representations, and this mapping is crucial for understanding the behavior of the perturbed system.
	
	We begin by recalling the relations between the creation and annihilation operators in the excitation basis:
	\begin{align}
		\hat{K}|k_{N_e},k_{N_e-1},\dots k_1\rangle &= \sum_{i=1}^{N_e}(1-\delta_{k_i,k_{i+1}})|p^{(i)}_{N_e},p^{(i)}_{N_e-1},\dots,p^{(i)}_1\rangle, \\
		\hat{K}^\dagger|k_{N_e},k_{N_e-1},\dots k_1\rangle &= \sum_{i=1}^{N_e+1}(1-\delta_{k_{i-1},k_{i}})|q^{(i)}_{N_e+1},q^{(i)}_{N_e},\dots q^{(i)}_1\rangle,
	\end{align}
	where $k_{N_e+1}=0$, $p_j^{(i)}=k_j-\delta_{i,j}$, $q_j^{(i)}=k_j+\delta_{i,j}$, and we formally require $\delta_{k_0,k_1}=0$. The Kronecker delta functions $\delta_{i,j}$ enforce the Pauli exclusion principle. These expressions describe how the action of $\hat{K}$ and $\hat{K}^\dagger$ modifies the excitation basis states.
	
	Next, we expand $(\hat{K}^\dagger)^n|0\rangle$ onto the excitation state basis:
	\[
	(\hat{K}^\dagger)^n |0\rangle = \sum_k b_{n,k} |\phi_n^k\rangle,
	\]
	where $b_{n,k}$ are the expansion coefficients, and $|\phi_n^k\rangle=|k_{N_e},k_{N_e-1}\dots k_1\rangle$ with the additional condition $\sum_{i=1}^{N_e} k_i = n$. This expansion allows us to identify the coefficients $b_{n,k}$, which are critical for understanding the underlying group structure of the excitations.
	
	The key idea is that each excitation basis state $|\phi_n^k\rangle$ can be mapped to an irreducible representation (irrep) of the permutation group.
	
\begin{figure*}[tbp]
		\includegraphics[width=0.8\textwidth]{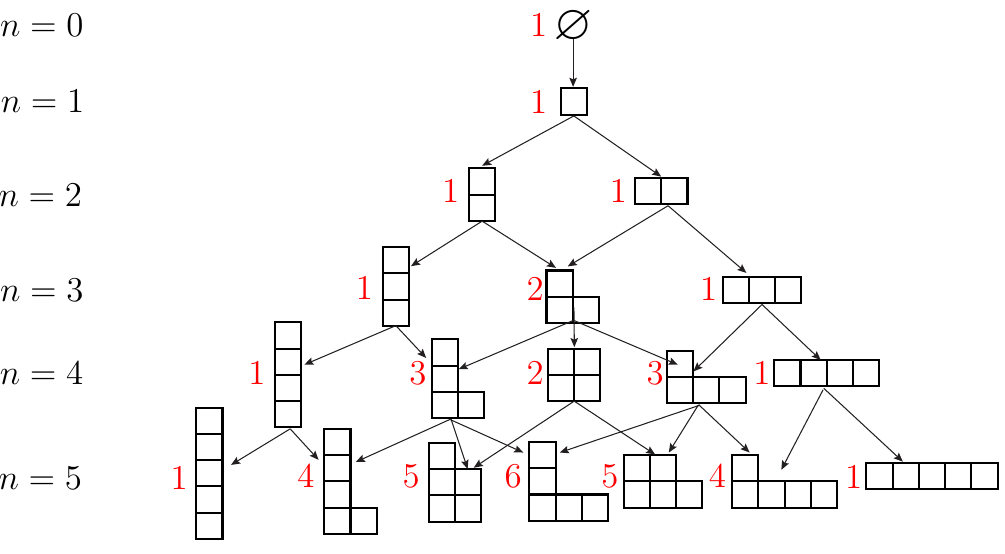}
		\caption{Young's lattice, showing the irreducible representations of the permutation group for different $n$. The numbers in red indicate the dimensions of each irreducible representation of the corresponding Young diagram. The arrows indicate the induction of the corresponding representation.}
		\label{Fig8}
	\end{figure*}

	The permutation group considers the permutation of $n$ individual numbers in an array. The so-called cycle notation $(a_1,a_2,\dots,a_n)$ describes the permutation of the elements from the positions $a_1,a_2,a_3,\dots,a_n$ to new positions $a_n,a_1,a_2,\dots,a_{n-1}$. Any permutation can be denoted using multiples of the cycle notation. For example, a permutation from ``123456'' to ``653124'' can be denoted as $(3)(25)(146)$.
	Consider an irreducible representation of the permutation group for a set of $n$ numbers.
	The representation is indexed by a partition of $n$. Here different partitions indicate different ways to perform the permutation.
	For example, with $n=6$, the partition $[1,2,3]$ indicates the permutation with the cycle notation $(a_1)(a_2,a_3)(a_4,a_5,a_6)$. In the literature, the Young diagram is often used to denote the irreducible representation.
	For an irreducible representation with the partition $[\phi_n^k]=[k_{N_e},k_{N_e-1}\dots k_1]$ satisfying $k_1\geq k_2\geq\dots\geq k_{N_e}$ and $\sum_jk_j=n$, the Young diagram consists of
	a total of $n$ layered boxes, with $k_j$ boxes in the $j$th layer (counting from the bottom layer). See Fig.~\ref{Fig8} for an illustration.

	The action of $\hat{K}$ and $\hat{K}^\dagger$ corresponds to the restriction and induction of the irreducible representation. This mapping is formalized by the following relations:
	\[
	\hat{K}(\hat{K}^\dagger)^n|0\rangle = n(\hat{K}^\dagger)^{n-1}|0\rangle,
	\]
	where the right-hand side follows from the structure of the induced and restricted representations of the permutation group. This relation holds for any dark state $|D\rangle$ of $\hat{K}$.
	
	To further elucidate this, we consider the partition $[k_{N_e}, k_{N_e-1}, \dots k_1]$, which indexes the irreducible representation. The structure of the Young diagram for this partition provides a visual way to understand how the excitation basis states map to the permutation group representations. The dimensions of these irreducible representations correspond to the expansion coefficients $b_{n,k}$ in the excitation basis, as shown in Fig.~\ref{Fig8}.
	
\begin{figure*}[tbp]
		\includegraphics[width=1\textwidth]{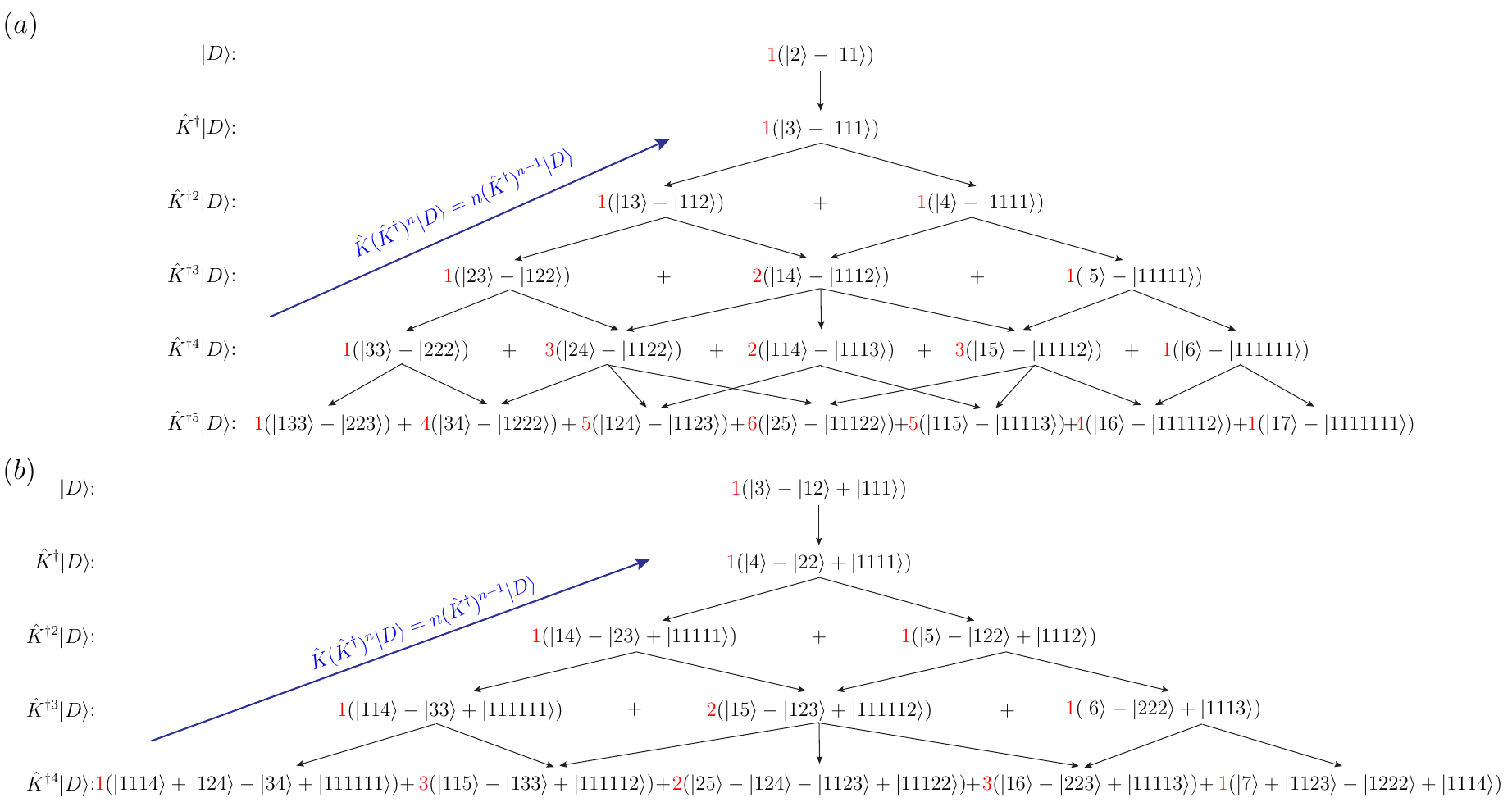}
		\caption{Excitations generated by $\hat{K}^\dag$ from different dark states. (a) The initial dark state is $|2\rangle-|11\rangle$. (b) The initial dark state is $|3\rangle-|12\rangle+|111\rangle$.}
		\label{Fig9}
	\end{figure*}

	We also find that the recurrence relations between the coefficients $b_{n,k}$ are governed by the branching rules of the permutation group. Specifically, the dimension $c_{n,k}$ of the irreducible representation can be computed through the branching rules, which relate the coefficients at different levels of the excitation hierarchy:
	\[
	c_{n,k} = \sum_{r \nearrow k} c_{n-1,r}, \quad n c_{n-1,r} = \sum_{r \nearrow k} c_{n,k}.
	\]
	These relations hold for all $n$, and they help to ensure that the coefficients $b_{n,k}$ obey the same rules as the irreducible representations of the permutation group.
	In Fig.~\ref{Fig9}(a) and (b), we show the excitations generated by $\hat{K}^\dagger$ from the dark states $|D\rangle=|2\rangle-|11\rangle$ and $|D\rangle=|3\rangle-|12\rangle+|111\rangle$, respectively. In both cases, $(\hat{K}^\dagger)^n|D\rangle$ can be expanded in the excitation basis, which can be mapped to the irreducible representations of the permutation group. The expansion coefficients are identical to the corresponding dimensions of the irreducible representations. We therefore surmise that, for any dark state $|D\rangle$ of $\hat{K}$, the relation $\hat{K}(\hat{K}^\dagger)^n|D\rangle = n(\hat{K}^\dagger)^{n-1}|D\rangle$ continues to hold. This conjecture is consistent with our numerical results throughout the work.

	Finally, we discuss how the finite system size modifies the behavior of the eigenmodes. For large enough $n$, the mapping between the excitation states and the permutation group representations remains valid, but at finite $L$ and $N$, the system?s limited size introduces a cutoff. This leads to a truncation of the perturbative expansion, and the series for the steady-state density matrix $\rho$ becomes approximate. The cutoff is determined by the number of fermion excitations that can be sustained on the lattice.
	
	For the steady state $\rho_0$ of $\mathcal{L}_0$, we define $l = \min(l^{|D^{B_1}\rangle}, l^{|D^{B_2}\rangle})$ as the maximum number of excitations allowed. The truncated perturbative series $\rho = \rho_0 + \sum_{n=1}^{l}s^n \rho_n$ is then an approximate eigenmode of $\mathcal{L}$, with the approximate eigenvalue $\lambda_f + i\Delta f \sim s^p$, where $p = 2l + 2$.
	
	This analysis provides a deep connection between the time crystal dynamics and the symmetry properties of the system, particularly in the context of the permutation-group representation. It is this connection that underlies the robustness of the prethermal time crystal, even in the presence of symmetry-breaking perturbations.
	
		\section{Liouvillian perturbation approach}
		In this Appendix, we outline the Liouvillain perturbation approach that leads to Eq.~(\ref{eq:recur}) . Following Ref.~\cite{Cavityskin}, we set $\mathcal{L}=\mathcal{L}_0+s\mathcal{L}_1$, with
		\begin{align}
			\mathcal{L}_0[\rho]&=-i[\Delta\hat{B}+ \gamma \hat{K}^\dagger\hat{K},\rho]+\gamma(2\hat{K}\rho\hat{K}^\dagger-\hat{K}^\dagger\hat{K}\rho-\rho\hat{K}^\dagger\hat{K}),\\
			\mathcal{L}_1[\rho]&=-i[\hat{K}+\hat{K}^\dagger,\rho].
		\end{align}
		For a non-dissipative eigenmode $\rho_0=|D^{B_1}\rangle\langle D^{B_2}|$ with $\mathcal{L}_0[\rho_0]=-i\Delta f\rho_0$, we set
		\begin{align}
			\rho_n=\sum_{j=0}^{n}\alpha_n^{(j)}\rho_n^{(j)},\quad\text{with}\quad \rho_n^{(j)}=\hat{K}^{\dagger j}\rho_0\hat{K}^{n-j},
		\end{align}
		with the coefficients
		\begin{equation}
	\alpha_n^{(j)}=\frac{i^{n-2j}s^n}{((1-i)\gamma-i\Delta)^{n-j}((1+i)\gamma+i\Delta)^j(n-j)!j!}.
	\label{eq:alphan}
\end{equation}
We therefore have
\begin{align}
	\mathcal{L}_0[\rho_{n}]+i\Delta f\rho_{n}&=
	-\sum_{j=0}^n \left((i\gamma+i\Delta)(2j-n)+\gamma n\right)\alpha_n^{(j)}\rho_n^{(j)} \\
	&\quad +2\gamma\sum_{j=1}^{n-1} j(n-j)\alpha_{n}^{(j)}\rho_{n-2}^{(j-1)}\label{L0rho},\\
	-s\mathcal{L}_1[\rho_{n-1}]&=
	is\sum_{j=1}^{n} \alpha_{n-1}^{(j-1)}\rho_n^{(j)}-is\sum_{j=0}^{n-1} \alpha_{n-1}^{(j)}\rho_n^{(j)} \\
	&\quad +is\sum_{j=1}^{n-1}\left(j\alpha_{n-1}^{(j)}-(n-j)\alpha_{n-1}^{(j-1)}\right)\rho_{n-2}^{(j-1)}.\label{L1rho}
\end{align}
		Note that Eqs.~(\ref{L0rho}) and (\ref{L1rho}) hold for $n\geq 2$. For $n=1$, the two equations still hold if we set the second term on the right of each to be zero.
		Plugging Eq.~(\ref{eq:alphan}) into Eqs.~(\ref{L0rho}) and (\ref{L1rho}), we have
		\begin{equation}
			\mathcal{L}_0[\rho_{n}]+i\Delta f\rho_{n}=-s\mathcal{L}_1[\rho_{n-1}],
		\end{equation}
		holding for $n\in\mathbb{N}^+$. Therefore, $\rho=\rho_0+\sum_{n=1}^\infty s^n\rho_n$ is a non-dissipative eigenmode of $\mathcal{L}$ with $\mathcal{L}[\rho]=-i\Delta f\rho$.
		However, when the lattice length $L$ or the atom number $N$ is finite, the relation $\hat{K}(\hat{K}^\dagger)^n|D^{B_{1,2}}\rangle=n(\hat{K}^\dagger)^{(n-1)}|D^{B_{1,2}}\rangle$ only holds up to a certain order $n\leq l$.
		This is because the excitation basis $|k_{N_e},k_{N_e-1},\dots k_1\rangle$ must satisfy $N_e\leq N$ and $k_1\leq L-N$ for finite $L$ or $N$. However, terms with
		$q_{N_e+1}^{(i)}=1$ and $q_1^{(i)}=k_1+1$ appear in the expression of  $\hat{K}^\dagger|k_{N_e},k_{N_e-1},\dots k_1\rangle$.
		To make the expression of
		$\hat{K}^\dagger|k_{N_e},k_{N_e-1},\dots k_1\rangle$ consistent with Eq.~(\ref{KDFock}),
		we must have $N_e\leq N-1$ and $k_1\leq L-N-1$.
		Similarly, to make the expression of $(\hat{K}^{\dagger})^{n}|k_{N_e},k_{N_e-1},\dots k_1\rangle$ consistent with Eq.~(\ref{KDFock}), we must have $n\leq N-N_e$ and $n\leq L-N-k_1$.
		Here $N-N_e$ indicates the number of consecutively occupied sites from the left (the real-space Fermi sea), and $L-N-k_1$ indicates the number of consecutively unoccupied sites from the right.
		In this context, since any dark state $|D\rangle$ can be expanded onto the excitation basis,
		we define $l_N^{|D\rangle}=\max\{j\big|{\langle D|\hat{c}_{j}^\dagger \hat{c}_{j}|D\rangle=1}\}$, labeling the site index to the left of which the Fermi sea is intact (without hole excitations).
		We also define $l_L^{|D\rangle}=\min\{j\big|{\langle D|\hat{c}_{L-j}^\dagger \hat{c}_{L-j}|D\rangle>0}\}$,
		labeling the site index to the right of which there are no fermion excitations.
		We then find that the relation $\hat{K}(\hat{K}^\dag)^{n}|D\rangle=n(\hat{K}^\dag)^{n-1}|D\rangle$ holds only up to $n\leq l^{|D\rangle}=\min(l_N^{|D\rangle},l_L^{|D\rangle})$.
		It follows that the expression $\mathcal{L}_0[\rho_{n}]+i\Delta f\rho_{n}=-s\mathcal{L}_1[\rho_{n-1}]$ also only holds for $n\leq l$, where $l=\min(l^{|D^{B_1}\rangle},l^{|D^{B_2}\rangle})$ (note that $\rho_0=|D^{B_1}\rangle\langle D^{B_2}|$).
		Then, the constructed density matrix $\rho=\rho_0+\sum_{n=1}^l s^n\rho_{n}$ can be regarded as an approximate eigenmode of $\mathcal{L}$, with the approximate eigenvalue calculated using $\mathrm{Tr}(\rho\mathcal{L}[\rho])$.
		Starting from the expression $\mathcal{L}[\rho]=-i\Delta f\rho+s^{l+1}\mathcal{L}_1[\rho_l]$, we have
		\begin{align}
			\mathrm{Tr}(\rho\mathcal{L}[\rho])+i\Delta f=\sum_{n=0}^{l}s^{l+n+1}\mathrm{Tr}(\rho_n\mathcal{L}_1[\rho_l]).
		\end{align}
		Now for each $\rho_0=|D^{B_1}\rangle\langle D^{B_2}|$ with $B_1 \ne B_2$, we have $\mathrm{Tr}(\rho_n^{(j)}\rho_m^{(k)})=0$.
		This is because $\mathrm{Tr}(\rho_n^{(j)}\rho_m^{(k)})=\mathrm{Tr}(\hat{K}^{\dag j}|D^{B_1}\rangle\langle D^{B_2}|\hat{K}^{(n-j)} \hat{K}^{\dag k} |D^{B_1}\rangle\langle D^{B_2}|\hat{K}^{(m-k)})$. And when
		$n-j\leq k$, we have $\langle D^{B_2}|\hat{K}^{(n-j)} \hat{K}^{\dag k} |D^{B_1}\rangle\propto \langle D^{B_2}|\hat{K}^{k+j-n} |D^{B_1}\rangle=0$, whereas for $n-j>k$, we have $\langle D^{B_2}|\hat{K}^{(n-j)} \hat{K}^{\dag k} |D^{B_1}\rangle\propto \langle D^{B_2}|\hat{K}^{\dag(n-k-j)} |D^{B_1}\rangle=0$.
		From $\mathrm{Tr}(\rho_n^{(j)}\rho_m^{(k)})=0$, we then have $\mathrm{Tr}(\rho_n\mathcal{L}_1[\rho_l])=0$ for $n\leq l$. Therefore, $\mathrm{Tr}(\rho\mathcal{L}[\rho])+i\Delta f$ is at least of the order $p=2l+2$ in $s$. The resulting expression $\lambda_f+i\Delta f\sim s^{2l+2}$ is consistent with the numerical results in Fig.~\ref{Fig3}(c).
		
	\end{appendix}

\end{document}